\input harvmac
\noblackbox
\def\npb#1#2#3{{\it Nucl.\ Phys.} {\bf B#1} (19#2) #3}

\def\plb#1#2#3{{\it Phys.\ Lett.} {\bf B#1} (19#2) #3}
\def\prl#1#2#3{{\it Phys.\ Rev.\ Lett.} {\bf #1} (19#2) #3}

\def\prd#1#2#3{{\it Phys.\ Rev.} {\bf D#1} (19#2) #3}

\def\ijmp#1#2#3{{\it Int.\ J. Mod.\ Phys.} {\bf A#1} (19#2) #3}
\def\cmp#1#2#3{{\it Commun.\ Math.\ Phys.} {\bf #1} (19#2) #3}

\def\jmp#1#2#3{{\it J. Math.\ Phys.} {\bf #1} (19#2) #3}
\def\cqg#1#2#3{{\it Class.\ Q. Grav.} {\bf #1} (19#2) #3}
\def\jpa#1#2#3{{\it J. Phys.} {\bf A#1} (19#2) #3}
\def\aop#1#2#3{{\it Ann.\ of Phys.} {\bf #1} (19#2) #3}

\def\endli{\hfill\break}
\def\frac#1#2{{#1 \over #2}}

\def\semi{\subset\kern-1em\times\;}
\def\bar#1{\overline{#1}}

\def\CA{{\cal A}}                   
\def\CD{{\cal D}}                   \def\CG{{\cal G}}
                   \def\CI{{\cal I}}
\def\CJ{{\cal J}}                   \def\CL{{\cal L}}
\def\CM{{\cal M}}                   
\def\CO{{\cal O}}                   \def\CP{{\cal P}}
\def\CR{{\cal R}}                   
\def\CV{{\cal V}}

                     \def\Z{{\bf Z}}
\def\osp{OSp\,}
\Title{\vbox{\baselineskip12pt
\hbox{hep-th/9712130}
\hbox{CALT-68-2152}}}
{\centerline{M-Theory as a Holographic Field Theory}}
\medskip\bigskip
\centerline{Petr Ho\v rava}
\medskip
\centerline{\it California Institute of Technology, Pasadena, CA 91125, USA}
\centerline{\tt horava@theory.caltech.edu}
\baselineskip18pt
\medskip\bigskip\medskip\bigskip\medskip
\noindent\baselineskip16pt
We suggest that M-theory could be non-perturbatively equivalent to a local 
quantum field theory.  More precisely, we present a ``renormalizable'' gauge 
theory in eleven dimensions, and show that it exhibits various properties 
expected of quantum M-theory, most notably the holographic principle of 
't~Hooft and Susskind.  The theory also satisfies Mach's principle: 
A macroscopically large space-time (and the inertia of low-energy excitations) 
is generated by a large number of ``partons'' in the microscopic theory.  We 
argue that at low energies in large eleven dimensions, the theory should be 
effectively described by eleven-dimensional supergravity.  This effective 
description breaks down at much lower energies than naively expected, 
precisely when the system saturates the Bekenstein bound on energy 
density.  We show that the number of partons scales like the area of the 
surface surrounding the system, and discuss how this holographic reduction of 
degrees of freedom affects the cosmological constant problem.  We propose the 
holographic field theory as a candidate for a covariant, non-perturbative  
formulation of quantum M-theory.  

\Date{December 1997}
\nref\cjs{E. Cremmer, B. Julia and J. Scherk, ``Supergravity Theory in 11 
Dimensions,'' \plb{76}{78}{409}.}
\nref\oldmthry{C.M. Hull and P.K. Townsend, ``Unity of Superstring 
Dualities,'' \npb{438}{95}{109}, hep-th/9410167.\endli
P.K. Townsend, ``The Eleven-Dimensional Supermembrane Revisited,'' 
\plb{380}{95}{184}, hep-th/9501068.\endli
E. Witten, ``String Theory Dynamics in Various Dimensions,'' 
\npb{443}{95}{85}, hep-th/9503124.}
\nref\power{J.H. Schwarz, ``The Power of M-Theory,'' \plb{367}{96}{97}, 
hep-th/9510086.}
\nref\hw{P. Ho\v rava and E. Witten, ``Heterotic and Type I String Dynamics 
from Eleven Dimensions,'' \npb{460}{96}{506}, hep-th/9510209.}
\nref\pktfour{P.K. Townsend, ``Four Lectures on M-Theory,'' hep-th/9612121.}
\nref\hweff{P. Ho\v rava and E. Witten, ``Eleven-Dimensional Supergravity on 
a Manifold with Boundary,'' \npb{475}{96}{94}, hep-th/9603142.}
\nref\bekenstein{J.D. Bekenstein, ``Entropy Bounds and Black Hole Remnants,'' 
\prd{49}{94}{1912}, gr-qc/9307035.}
\nref\tholog{G. 't~Hooft, ``Dimensional Reduction in Quantum Gravity,'' 
gr-qc/9310026, in: ``Salamfestschrift,'' eds: A.~Ali, J.~Ellis, 
S.~Randjbar-Daemi (World Scientific, 1993).}
\nref\sholog{L. Susskind, ``The World as a Hologram,''\jmp{36}{95}{6377}, 
hep-th/9409089.}
\nref\threv{G. 't~Hooft, ``Quantum Information and Information Loss in General 
Relativity,'' gr-qc/9509050; ``The Scattering Matrix Approach for the Quantum 
Black Hole: An Overview,'' gr-qc/9607022.}
\nref\dshort{U.H. Danielsson, G. Ferretti and B. Sundborg, ``D-Particle 
Dynamics and Bound States,''  \ijmp{11}{96}{5463}, hep-th/9603081.\endli
D. Kabat and P. Pouliot, ``A Comment on Zero-Brane Quantum Mechanics,'' 
\prl{77}{96}{1004}, hep-th/9603127.\endli
M.R. Douglas, D. Kabat, P. Pouliot and S.H. Shenker, ``D-Branes and Short 
Distances in String Theory,'' \npb{485}{97}{85}, hep-th/9608024.}
\nref\bfss{T. Banks, W. Fischler, S.H. Shenker and L. Susskind, ``M Theory as 
a Matrix Model: A Conjecture,'' \prd{55}{97}{5112}, hep-th/9610043.}
\nref\sdlcq{L. Susskind, ``Another Conjecture about M(atrix) Theory,'' 
hep-th/9704080.}
\nref\mtss{A. Sen, ``D0-Branes on $T^n$ and Matrix Theory,'' hep-th/9709220.
\endli
N. Seiberg, ``Why is the Matrix Model Correct?'' hep-th/9710009.}
\nref\matrev{T. Banks, ``Matrix Theory,'' hep-th/9710213.\endli
D. Bigatti and L. Susskind, ``Review of Matrix Theory,'' hep-th/9712072.}
\nref\gsw{M.B. Green, J.H. Schwarz and E. Witten, ``Superstring Theory,'' 
Vol.~1 and~2 (Cambridge University Press, 1987).}
\nref\ewcs{E. Witten, ``Quantum Field Theory and the Jones Polynomial,'' 
\cmp{121}{89}{351}.}
\nref\cssgr{P. van Nieuwenhuizen, ``Three-Dimensional Conformal Supergravity 
and Chern-Simons Terms,'' \prd{32}{85}{872}.\endli
A. Ach\'ucarro and P.K. Townsend, ``A Chern-Simons Action for 
Three-Dimensional Anti-de~Sitter Supergravity Theories,'' \plb{180}{86}{89}.}
\nref\ewcsg{E. Witten, ``$2+1$ Dimensional Gravity as an Exactly Soluble 
System,'' \npb{311}{88}{46}, ``Topology-Changing Amplitudes in $2+1$ 
Dimensional Gravity,'' \npb{323}{89}{113}.}
\nref\alicham{A.H. Chamseddine, ``Topological Gauge Theory of Gravity in Five 
and All Odd Dimensions,'' \plb{233}{89}{291}; ``Topological Gravity and 
Supergravity in Various Dimensions,'' \npb{346}{90}{213}.}
\nref\csbh{M. Ba\~nados, C. Teitelboim and J. Zanelli, ``The Black Hole in 
Three Dimensional Space Time,'' \prl{69}{92}{1849}, hep-th/9204099; 
``Dimensionally Continued Black Holes,'' \prd{49}{94}{975}, gr-qc/9307033; 
``Black Hole Entropy and the Dimensional Continuation of the Gauss-Bonnet  
Theorem,'' \prl{72}{94}{957}.}
\nref\cssuper{M. Ba\~nados, R. Troncoso and J. Zanelli, ``Higher-Dimensional 
Chern-Simons Supergravity,'' \prd{54}{96}{2605}, gr-qc/9601003.\endli
R. Troncoso and J. Zanelli, ``New Gauge Supergravity in Seven and Eleven 
Dimensions,'' hep-th/9710180.}
\nref\machb{``Mach's Principle: {}From Newton's Bucket to Quantum Gravity,'' 
eds: J.B. Barbour and H. Pfister (Birkh\"auser, Boston, 1995).}
\nref\aibook{J.A. de Azc\'arraga and J.M. Izquierdo, ``Lie Groups, Lie 
Algebras, Cohomology and Some Applications in Physics'' (Cambridge University 
Press, 1995).}
\nref\zanelli{J. Zanelli, ``Quantization of the Gravitational Constant in 
Odd-Dimensional Gravity,'' \prd{51}{95}{490}, hep-th/9406202.}
\nref\ewflux{E. Witten, ``On Flux Quantization in $M$-Theory and the Effective 
Action,'' hep-th/9609122.}
\nref\vhvp{J.W. van~Holten and A. Van~Proeyen, ``$N=1$ Supersymmetry Algebras 
in $d=2$, 3, 4 mod 8,'' \jpa{15}{82}{3763}.}
\nref\hidden{R. D'Auria and P. Fr\'e, ``Geometric Supergravity in $D=11$ and 
Its Hidden Supergroup,'' \npb{201}{1982}{101}.}
\nref\pkt{P.K. Townsend, ``$p$-Brane Democracy,'' hep-th/9507048; ``M-Theory 
from its Superalgebra,'' hep-th/9712004.}
\nref\bars{I. Bars, ``S-Theory,'' \prd{55}{97}{2373}, hep-th/9607112; 
``A Case for 14 Dimensions,'' \plb{403}{97}{257}, hep-th/9704154.}
\nref\csgtbound{S. Elizur, G. Moore, A. Schwimmer and N. Seiberg, ``Remarks on 
the Canonical Quantization of the Chern-Simons-Witten Theory,'' 
\npb{326}{89}{108}.}
\nref\carlip{S. Carlip, ``Exact Quantum Scattering in $2+1$-Dimensional 
Gravity,'' \npb{324}{89}{106}.}
\nref\hallcs{A. Zee, ``Quantum Hall Fluids,'' cond-mat/9501022.\endli
A. Lopez and E. Fradkin, ``Fermionic Chern-Simons Field Theory for the 
Fractional Quantum Hall Effect,'' cond-mat/9704055.}
\nref\condmat{F. Wilczek, ``Fractional Statistics and Anyon 
Superconductivity'' (World Scientific, 1990).\endli
E. Fradkin, ``Field Theories of Condensed Matter Systems'' (Addison-Wesley, 
1991).}
\nref\morehidden{ L.~Castellani, P.~Fr\'e, F.~Giani, K.~Pilch and 
P.~van~Nieuwenhuizen, ``Gauging of $d=11$ Supergravity?'' \aop{146}{83}{35}.
\endli
R.~D'Auria, P.~Fr\'e, P.K.~Townsend and P.~van~Nieuwenhuizen, ``Invariance of 
Actions, Rheonomy, and the New Minimal $N=1$ Supergravity in the Group 
Manifold Approach,'' \aop{155}{84}{423}.\endli
P. Fr\'e, ``Comments on the 6-Index Photon in $d=11$ Supergravity 
and the Gauging of Free Differential Algebras,'' \cqg{1}{84}{L81}.}
\nref\sugrgeo{L. Castellani, R. D'Auria and P. Fr\'e, ``Supergravity and 
Superstrings: A Geometric Perspective,'' Vols.\ 1-3 (World Scientific, 
1991).}
\nref\uniquesg{K. Bautier, S. Deser, M. Henneaux and D. Seminara, ``No 
Cosmological $D=11$ Supergravity,'' \plb{406}{97}{49}.}
\nref\banks{T. Banks, ``SUSY Breaking, Cosmology, Vacuum Selection and the 
Cosmological Constant in String Theory,'' hep-th/9601151.}
\nref\weinberg{S. Weinberg, ``The Cosmological Constant Problem,'' {\it 
Rev.\ Mod.\ Phys.} {\bf 61} (1989) 1.}
\nref\cbound{S. Carlip, ``The Statistical Mechanics of the 
Three-Dimensional Euclidean Black Hole,'' \prd{55}{97}{878}, gr-qc/9606043.}
\nref\bala{A.P. Balachandran, L. Chandar and E. Ercolessi, ``Edge 
States in Gauge Theories: Theory, Interpretations and Predictions,'' 
\ijmp{10}{95}{1969}, hep-th/9411164.\endli
A.P. Balachandran, L. Chandar and Arshad Momen, ``Edge States in Gravity and 
Black Hole Physics,'' \npb{461}{96}{581}, gr-qc/9412019.}
\newsec{Introduction}

M-theory has emerged from our understanding of non-perturbative string 
dynamics, as a hypothetical quantum theory which has eleven-dimensional 
supergravity \cjs\ as its low-energy limit, and is related to string theory 
via various dualities \refs{\oldmthry-\hw} (for an introduction and 
references, see e.g.\ \pktfour ).  While the low-energy effective description 
of this theory in terms of eleven-dimensional supegravity (coupled to $E_8$ 
Yang-Mills supermultiplets if the space-time manifold has boundaries 
\refs{\hw,\hweff}) is relatively well understood, we still need to clarify 
how M-theory is to be formulated as a non-perturbative quantum theory.  

Our search for a non-perturbative formulation of quantum M-theory can be 
guided by some general observations.  First of all, M-theory should represent, 
among other things, a consistent quantum theory of gravity.  Using the 
Bekenstein bound on the maximum entropy in a given region of space 
\bekenstein , 't~Hooft and Susskind have argued very convincingly that any 
such theory should satisfy the holographic principle \refs{\tholog,\sholog} 
(see also \threv ).  The holographic property predicts a radical reduction of 
the number of degrees of freedom in quantum theory of gravity; unlike in any 
conventional local field theory, their number should scale like the area 
surrounding the system.   

Other observations come from our improved understanding of non-perturbative 
string theory.  At sub-stringy distances, a new regime of weakly coupled 
string theory has been discovered and analyzed \dshort .  In this regime, the 
short distance physics is dominated by D0-branes, and long-distance gravity is 
replaced by Yang-Mills gauge theory on world-volumes of branes.  The matrix 
theory formulation of quantum M-theory, proposed by Banks, Fischler, 
Shenker and Susskind \refs{\bfss,\sdlcq}, takes this lesson very seriously 
and elevates some of the crucial features of D-branes to eleven dimensions, 
using a light-front formulation of M-theory. Sen and Seiberg have recently 
presented an elegant heuristic scaling argument \mtss , which provides a 
rationale for the matrix theory proposal and clarifies it significantly.  

Matrix theory has proven to be a very impressive candidate for the 
non-perturbative formulation of M-theory.  (For recent reviews, see 
\matrev .)  Despite its outstanding successes, however, it still leaves many 
important questions unanswered.  It is background-dependent and non-covariant, 
and the scaling arguments of \mtss\ suggest the existence of conceptual 
problems for compactifications on tori of dimension higher than five.  

In general, it has been suggested that since M-theory cannot be a string 
theory, it must be a new kind of theory, which should perhaps be formulated in 
terms of completely new degrees of freedom, and require new physical 
principles.  This may even lead to a change in our way of thinking about 
microscopic physics, perhaps as radical as the discovery of quantum 
mechanics.  (Indeed, a certain amusing analogy between the development 
of quantum mechanics and that of string theory has been pointed out, see 
\refs{\gsw,\ page\ 1}.)  

We would like to point out a different analogy, which relates the current 
situation in M-theory to the situation in the theory of strong interactions 
before the discovery of QCD.  In the mid-60's, it was generally believed that 
in order to understand strong interactions, local quantum field theory would 
have to be abandoned altogether, and radically new physical principles would 
be needed.  The efforts to go beyond field theory indeed initiated the 
development of important new concepts, such as the analytic S-matrix, 
bootstrap, duality, Regge trajectories etc.  However, we know that at the end, 
the puzzle of the theory of strong interactions turned out to have a beautiful 
resolution in the ``conservative'' framework of local quantum field theory.  

In this paper we will adopt a similarly ``conservative'' approach to 
M-theory.  Instead of looking for radically new principles and degrees of 
freedom, we will present evidence suggesting that M-theory may in fact be 
equivalent to a local quantum field theory.  

Our starting point in Section~2 will be a Yang-Mills gauge theory in eleven 
dimensions.  The gauge group is a certain supersymmetric extension of the 
eleven-dimensional anti-de~Sitter group, but the theory should not be confused 
with de~Sitter supergravity.  Microscopically, our theory is a gauge theory, 
with Yang-Mills gauge symmetries.  In addition to the gauge symmetries, we 
require invariance under space-time diffeomorphisms, as well as parity 
invariance.  

All terms in the Lagrangian that are allowed by the symmetries are of higher 
order in fields, and are in fact given by Chern-Simons terms.  Thus, our 
theory belongs to the class of Chern-Simons gauge theories \ewcs .  
Chern-Simons gravity was first studied in $2+1$ dimensions 
\refs{\cssgr,\ewcsg}, and then extended to higher odd dimensions 
\refs{\alicham-\cssuper}.  Our formulation will closely follow that of 
\refs{\ewcsg,\alicham}.  

When expanded around maximally symmetric solutions, the theory has no 
propagator, and the low-energy field theory is ill-defined, or at least 
difficult to understand with conventional methods.  In Section~3, we adopt 
the following effective-theory approach to this 
issue.  We will not attempt to quantize the theory microscopically.  Instead, 
we will try to identify a low-energy regime, in which the theory does have a 
conventional low-energy effective field theory description, with excitations 
propagating in a macroscopically large space-time.  

In order to find such a macroscopic low-energy regime, we will have to 
introduce matter, in the form of first-quantized particles (or ``partons'') 
represented by Wilson lines -- the only objects that couple naturally to the 
gauge field.  A large space-time will require a large number of partons.  We 
will see that the theory satisfies Mach's principle \machb :  Macroscopically 
large space-times and the inertia of propagating low-energy degrees of freedom 
will be generated by the distribution of a large number of partons in the 
theory.  

In Section~4 we study the theory at low-energies in large eleven dimensions.  
We will show that for the appropriate choice of the gauge group, the flat 
eleven-dimensional space-time is a solution of the theory, in a mean field 
approximation which replaces the effect of individual partons at large 
distances by a uniform density of partons.  We identify the regime which has 
low-energy degrees of freedom propagating in a large space-time, and argue 
that the low-energy physics is effectively described by eleven-dimensional 
supergravity.  We demonstrate that this effective supergravity description 
naturally breaks down at energies much lower than the naively expected Planck 
scale.  

In Section~5 we show that the breakdown of the low-energy effective theory is 
in accord with the Bekenstein bound on energy density, and that the 
theory in fact satisfies the holographic principle.  More precisely, we 
demonstrate that the limit of validity of the low-energy effective 
supergravity description is reached precisely when the energy in any region of 
characteristic size $L$ equals the mass of the Schwarzschild black hole 
with radius $L$.  We will see that as expected in a holographic theory, the 
number of partons in the system scales as the area of the surface surrounding 
the system.  At large distances and low energies, the theory is described by 
low-energy supergravity, and space-time diffeomorphisms are a part of the 
dynamical gauge group.  The realization of the holographic principle in local 
field theory also sheds some new light on the cosmological constant problem, 
and we will find indications suggesting that $\Lambda$ may be naturally small 
in holographic field theory.  

The purpose of this paper is to stress some of the unexpected features of the 
theory, in particular the holographic property, and to set the ground for a 
more detailed investigation.  Our presentation will be rather sketchy, and 
we will leave out many details and open questions for further study.  

\newsec{The theory}

Consider a gauge field theory in eleven dimensions, defined as follows.  
Start with an eleven-dimensional manifold $\CM$, with coordinates $x^M$, $M=0, 
\ldots, 10$.  Our theory will be a gauge theory described by a Yang-Mills 
one-form potential $A$, in the adjoint representation of a certain gauge group 
$\CG$.  We impose gauge invariance under the Yang-Mills gauge transformations 
\eqn\eegsymm{\delta A_M=D_M\varepsilon.}
There is no preferred metric on $\CM$, and we require that the theory be 
invariant under local diffeomorphisms of $\CM$.  The only Lagrangian that 
respects these symmetries is the Chern-Simons Lagrangian 
\eqn\eelagr{\CL=-\frac{1}{g^2}\int_\CM\omega_{11}(A),}
where $\omega_{11}(A)$ is a Chern-Simons eleven-form, defined by
\eqn\eecsdef{d\omega_{11}(A)=\Tr\left(F\wedge\ldots\wedge F\right).}
Here $F$ is the field strength associated with $A$, ``Tr'' refers to a 
symmetric, invariant six-tensor on $\CG$.  In fact, the Lagrangian can be a 
linear combination of all possible such terms if there is more than one 
invariant six-tensor on $\CG$ that satisfy all other symmetry restrictions 
we may want to impose on the theory; each term would then have its own 
coupling constant $g$.  The theory is renormalizable in the elementary sense 
that all couplings allowed by the symmetries are marginal.  

Equation \eecsdef\ can be solved, leading to an explicit formula for the 
Chern-Simons form $\omega_{11}(A)$ (see e.g.\ \aibook ), 
\eqn\eecsexp{\omega_{11}(A)=6\int_0^1dt\;\Tr\left(A\wedge (tdA+t^2A\wedge A)
\wedge\ldots\wedge(tdA+t^2A\wedge A)\right).}
The leading term in $\omega_{11}(A)$ is proportional to $\Tr (A\wedge dA
\wedge\ldots\wedge dA)$; all other terms are of higher order in $A$.  

The coupling constant $g$ in \eelagr\ is dimensionless.  It may be quantized, 
depending on the precise choice of the gauge group and ``Tr''.  The 
quantization condition can be derived as follows.  Consider a 
twelve-dimensional manifold $\CB$ whose boundary is $\CM$, and extend the 
gauge connection over $\CB$.  The Lagrangian \eelagr\ is then more precisely 
defined using \eecsdef , as an integral of $\Tr (F\wedge\ldots F)$ over $\CB$: 
\eqn\eelagtwelve{\CL=-\frac{1}{g^2}\int_\CB\Tr\left(F\wedge\ldots\wedge F
\right).}
The quantization condition on the coupling arises from the requirement that 
$\CL$ be independent of $\CB$ and the way $A$ has been extended over $\CB$.  
Typically, this leads to 
\eqn\eeqcoup{\frac{1}{g^2}\sim k,}
with $k$ an integer.%
\foot{In the case of the de~Sitter gauge group, directly relevant to the 
present paper, the issue of coupling constant quantization has been discussed 
in \zanelli.}

So far we have imposed local diffeomorphism invariance as the only symmetry 
in addition to local gauge invariance.  Our understanding of low-energy 
effective M-theory indicates that any candidate for non-perturbative 
formulation of M-theory should also be parity invariant.%
\foot{We know that M-theory is parity invariant \refs{\hw,\ewflux}. Indeed, in 
M-theory parity can be gauged, leading to the sector of heterotic vacua of the 
theory.}
The $\Z_2$ transformation $\CP_0$ that changes space-time orientation by 
reversing one of the space-time dimensions (say $x^1$) cannot be a symmetry of 
the Chern-Simons gauge theory, since each Chern-Simons form is odd under 
$\CP_0$.  In order to become a symmetry, $\CP_0$ has to be accompanied by an 
involution $\CI$ on the gauge group $\CG$.  Depending on the choice of $\CG$ 
and $\CI$, the microscopic theory will be constrained by the requirement of 
invariance under parity, now defined as 
\eqn\eeimpparity{\CP=\CP_0\cdot\CI,}
leading to restrictions on admissible ``Tr'' that can appear in \eelagr .  

\subsec{Gauge group and parity invariance}

As our gauge group, we will choose a supersymmetric extension of the 
anti-de~Sitter group in eleven dimensions.  We need the de~Sitter group 
as a part of the microscopic gauge group, because only in that case we will 
eventually find a low-energy regime described by effective supergravity 
with the conventional Lagrangian linear in curvature, and the flat space as a 
solution of the low-energy theory.  

The anti-de~Sitter group is generated by $P_A$ and $J_{AB}$, with $A,B=0,
\ldots 10$. There is an invariant six-tensor on the anti-de~Sitter group that 
will play crucial role in our theory, 
\eqn\eeadstr{\left\langle P_AJ_{A_1A_2}\ldots J_{A_9A_{10}}\right\rangle
=\epsilon_{AA_1\ldots A_{10}}}
(with all other terms zero).  This six-tensor defines a Chern-Simons 
eleven-form of the anti-de~Sitter group.  Chern-Simons gravity with this 
Lagrangian was first studied in various dimensions by Chamseddine \alicham .  
Our Lagrangian will be a supersymmetric extension of this bosonic Chern-Simons 
Lagrangian.  

To make any contact with M-theory, we need at least 32 supercharges.  
It was shown by van~Holten and Van~Proeyen in \vhvp\ that the minimal 
supersymmetric extension of the eleven-dimensional anti-de~Sitter group into a 
supergroup with a 32-component supercharge $Q_\alpha$ requires the 
introduction of an extra bosonic five-form charge $K_{A_1\ldots A_5}$, which 
extends the group to $\osp(1|32)$.  

We want to impose parity invariance as a symmetry of our theory.  It turns out 
that the minimal supersymmetric extension $\osp(1|32)$ of the anti-de~Sitter 
group is not compatible with parity.  Indeed, we know how $\CI$ should act on 
the bosonic anti-de~Sitter generators: Both $P_A$ and $J_{AB}$ flip signs 
whenever $A$ or $B=1$.  On the fermionic generators, $\CI$ acts by
\eqn\eeiactferm{Q_\alpha\rightarrow (\Gamma_{1}Q)_\alpha.}
It is easy to see that $\CI$ cannot be extended to an automorphism of 
$\osp(1|32)$.  The obstruction comes from the higher-form sector of the 
algebra.  It is natural to extend $\CI$ to the five-form charge in such a 
way that it changes sign whenever $A_i=1$ for any $i=1,\ldots 5$.  
However, this rule does not respect the group structure of $\osp(1|32)$,  
roughly because of the presence of the antisymmetric $\epsilon$ 
tensor in some of the commutation relations.  

Thus, parity invariance will require a non-minimal extension of the 
anti-de~Sitter group, into a group with 64 supercharges.%
\foot{First indications that the symmetry algebra underlying 
eleven-dimensional supergravity may contain 64 supercharges appeared in 
\hidden .  The importance of algebraic structure in M-theory has been stressed 
by Townsend \pkt\ and Bars \bars .  Indeed, 64 supercharges appeared in this 
algebraic approach to M-theory \bars , as a part of the maximal supersymmetric 
algebra that could contain all string dualities.}
The minimal choice of the gauge group compatible with parity invariance will 
contain extra, higher-form bosonic charges $K_{A_1\ldots A_r}$ for some set of 
values of $r$, and an extra 32-component supercharge $Q'_\alpha$.  We can now 
extend the definition of $\CI$ to these new charges, requiring that the 
bosonic charges change sign under $\CI$ whenever either of their indices 
equals 1, and $Q'\rightarrow -\Gamma_1Q'$.  The minimal set of charges that 
allow commutation relations that respect this $\CI$ will contain a six-form, 
a nine-form, and a ten-form charge, in addition to $P_A$, $J_{AB}$ and 
$K_{A_1\ldots A_5}$.  (Heuristically, we need a dual charge for each of the 
original bosonic charges, in order to write down commutation relations without 
the antisymmetric $\epsilon$ tensor.)  These charges generate a group 
isomorphic $\osp(1|32)\times\osp(1|32)$, which happens to be the non-chiral 
super Lorentz group in twelve dimensions with signature (10,2) \vhvp .  
The bosonic charges form the Lie algebra of $Sp\,(32)\times Sp\,(32)$.  
(For details, see \vhvp .) 

We will parametrize the components of the gauge field $A$ in the adjoint of 
$\osp(1|32)\times\osp(1|32)$ as follows, 
\eqn\eegfconv{A_M=V_M^AP_A+\frac{1}{2}\omega_M^{AB}J_{AB}+\sum_r \frac{1}{r!}
B_M^{A_1\ldots A_r}K_{A_1\ldots A_r}+\psi_M^\alpha Q_\alpha+\eta_M^\alpha 
Q'_\alpha,}
where we have denoted all bosonic higher-form charges collectively by 
$K_{A_1\ldots A_r}$, with $r=5,6,9,10$.  

Our theory is formally defined by the path integral,
\eqn\eedefthy{\int\CD Ae^{i\CL}.}
We will mostly discuss classical aspects of the theory in this paper, and 
will not analyze the precise definition of the measure in \eedefthy .  
Our focus will be on an effective approach, and we will try to identify a 
regime in this microscopic theory where interesting low-energy physics appears 
already at tree level.  

Since the Lagrangian is of higher order in fields, this theory does not have 
a standard kinetic term; moreover, it is topological in the sense that no 
metric has been used to write down the theory.  Notice that the theory still 
has dynamical degrees of freedom, as the equations of motion are 
\eqn\eeeom{F\wedge F\wedge F\wedge F\wedge F=0.}
There is however no standard propagator for these local degrees of freedom in 
the $F=0$ vacuum, nor is there a conventional perturbation theory in terms of 
weakly coupled localized multi-particle states.  

\newsec{Large universes and Mach's principle}

We live in a large universe, whose behavior at low energies seems well 
described by a local quantum field theory of particle-like excitations.  We 
want to identify a regime in our theory, which has such a low-energy effective 
description.  In particular, we would like our theory to have an 
eleven-dimensional vacuum described at low energies by eleven-dimensional 
supergravity, with flat eleven-dimensional space-time as a solution.  

\subsec{Effective theory in a large universe}

First of all, we would like to write down the flat space-time as a particular 
gauge field configuration.  We want to identify the $P_A$ component of the 
gauge field with the vielbein field, and the $J_{AB}$ component with the spin 
connection.  However, the gauge field $A_M=V_M^AP_A+\omega_M^{AB}J_{AB}+
\ldots$ is of dimension one, while the natural dimension for the vielbein is 
zero.  We introduce the dimensionless vielbein $e_M^A$, and write 
\eqn\eegenmsp{V_M^A=M e_M^A.}
We will use $\bar e_M^A$ to denote the flat eleven-dimensional vielbein, 
$\bar e_M^A=\delta_M^A$.  Hence, the gauge field configuration that represents 
the flat eleven-dimensional space-time is 
\eqn\eeflatmsp{\bar A_M=M\bar e_M^AP_A.}
We were able to write down the flat space-time geometry as a particular 
gauge field $\bar A$, at the cost of introducing a mass scale $M$ into the 
theory.  This mass scale is not a part of the path integral definition of our 
theory.  Rather, it appears as a property of the particular gauge 
configuration $\bar A$. 

The mass scale $M$ can be interpreted as the inverse characteristic size of 
the universe (or, more generally, of the box large enough to contain our 
system).  Indeed, the ``dimensionless volume'' of a ten-dimensional space-like 
hypersurface $\CM_{10}\subset\CM$ 
\eqn\eedlessvol{\int_{\CM_{10}}V\wedge\ldots\wedge V}
is a number of order one, which gives for the standard volume 
\eqn\eedfulvol{\CV=\int_{\CM_{10}}e\wedge\ldots\wedge e\sim
\frac{1}{M^{10}}.}
Of course, this argument could be easily refined to include the case with a 
flat metric on $\CM_{10}$ of toroidal topology; the radii of the torus would 
then be measured in units of $L\equiv M^{-1}$.  

There are two puzzles that we have have to resolve in our scenario.  First, 
the flat eleven-dimensional space-time \eeflatmsp\ is not a solution of the 
classical equations of motion of our $\osp(1|32)\times\osp(1|32)$ Chern-Simons 
gauge theory.  There {\it is\/} a formal solution of the equation of motion, 
which looks like the anti-de~Sitter space.  However, there is no conventional 
low-energy effective theory that would result from expanding the microscopic 
gauge theory around the anti-de~Sitter solution.  In particular, the 
formal expansion would have no quadratic term in the Lagrangian, and no 
propagator for particle-like degrees of freedom.  According to the logic of 
our approach, we are only interested in low-energy regimes that have a 
conventional effective field theory description.  

Another puzzling feature of the theory is the presence of a dimensionless 
coupling $g$ in \eelagr .  We know that M-theory -- at least at low energies, 
where it is well described by eleven-dimensional supergravity -- does not 
contain any such free dimensionless parameters.  If our theory is to be a 
reasonable candidate for the microscopic description of M-theory, we have to 
explain why $g$ does not appear as a free dimensionless coupling in the 
theory at low energies.  

We will see momentarily how both of these issues are resolved when we 
introduce partonic matter into the theory.  The discrete coupling constant 
$k$ that appears in \eeqcoup\ will turn out to play the role of the number of 
elementary constituents (``partons'')  in our 
system.  Only for a large number of partons, our theory will have a low-energy 
description in terms of supergravity degrees of freedom propagating in a 
macroscopically large space-time.  This relation between the number of partons 
and the size of the low-energy world is a first indication that our theory 
satisfies Mach's principle.  

\subsec{Matter and Mach's principle}

The gauge field is a one-form, and it couples naturally to point particles.  
Consider the Wilson line 
\eqn\eewl{W_\CR (C)=\tr_\CR P\exp\int_CA,}
where $\CR$ is a representation of the gauge group, and $C$ is a certain 
contour in $\CM$.  The Wilson line defines an observable in our gauge theory, 
and one can study physical processes that involve correlation functions of a 
certain number of such Wilson lines.  This is in fact the most natural way of 
introducing matter in our theory.  The Wilson lines correspond to trajectories 
of particles of matter; their species are in correspondence with the 
representations of the gauge group.  These particles will play the role of 
``partons'' in our microscopic theory.  

Consider now a universe $\CM$ with $N$ Wilson lines (or ``partons'') inside.   
The partons couple to the gauge fields through their current $\CJ$, which is 
a sum of delta functions localized at their corresponding contours $C_i$.  
{}For $N$ Wilson lines the current is 
\eqn\eewlj{\CJ=\sum_{i=1}^Nj^aT_a\delta(C_i),}
(here $T_a$ collectively denotes all generators of the gauge group), 
and the Lagrangian in the presence of the Wilson lines is modified to 
\eqn\eelagwl{\CL=-\frac{1}{g^2}\int_\CM\omega_{11}(A)+\int_\CM \tr(A\wedge
\CJ).}
Notice that since the group generators $T_a$ in \eewlj\ are matrices in the 
representations $\CR_i$ of the gauge group, their presence in the Lagrangian 
needs further interpretation.  The $T_a$ in \eewlj\ should be properly 
interpreted as quantum objects that emerge from the quantization of extra 
degrees of freedom localized at the contours $C_i$.  This is of course a 
procedure standard in gauge theories in general, and in Chern-Simons theories 
in particular \refs{\ewcs,\csgtbound}, and we will not repeat the details 
here.  (See \refs{\ewcsg,\carlip} for more details on this construction in 
the case of $2+1$ Chern-Simons gravity.)  

The equations of motion in the presence of $N$ partons no longer require 
the wedge product of five $F$'s to vanish.  Rather, the flux of the gauge 
field is tied to the current:  
\eqn\eeeomwl{F\wedge\ldots\wedge F=g^2\CJ.}
Thus, the partons serve as sources for the field strength flux, which is 
non-zero and localized at the $N$ contours $C_i$, and zero outside the 
trajectories of the partons.  

In the next section, we will be interested in describing such system at 
large distances, where the collective effect of a large number of Wilson lines 
can be summarized in terms of a uniform mean field, representing macroscopic 
space-time geometry.  Our theory is actually an implementation of Mach's 
principle \machb :  The geometry of space-time is generated as a collective 
effect by the distribution of matter (represented by the partons) in the 
microscopic theory.  The flat macroscopic space-time emerges as a collective 
effect, in the presence of a non-trivial matter distribution.  In the absence 
of matter, not even an empty, flat macroscopic space-time is possible.  At low 
energies, our theory also satisfies Mach's principle in another of its classic 
formulations:  The inertia of propagating particle-like degrees of freedom is 
generated as a collective effect determined by the distribution of matter in 
the microscopic theory. 

\newsec{Low-energy effective supergravity in eleven dimensions}

\subsec{Mean field theory and flat eleven-dimensional space-time}

We are interested in the physics at distances much larger than the 
characteristic distance between two partons.  At those distances, we can 
effectively approximate the source $\CJ$ -- which is microscopically a sum of 
$N$ delta functions \eewlj\ -- by a uniform density field $J$, 
\eqn\eecurrmfa{J=cNM^{10}\epsilon_{A_1\ldots A_{11}}P^{A_1}\bar e^{A_2}
\wedge\ldots\wedge\bar e^{A_{11}}.}
We expect the mean field approximation to be valid at distances much larger 
than the characteristic distance between partons as defined a posteriori by 
$\bar e_M^A$.  This approximation is somewhat reminiscent of the average field 
approximation frequently used in the theory of condensed matter systems 
described by Chern-Simons theory; see e.g.\ \refs{\hallcs,\condmat}.  

We will adopt this mean field ansatz for the rest of the paper, and will not
attempt to derive it from the microscopic theory.  In particular, we will not 
identify precisely the species of partons that leads to the mean field 
current, leaving this very important point to future study.  

In order to write down the mean field ansatz \eecurrmfa\ for $\CJ$ in terms 
of the flat space vielbein $\bar e_M^A$, we had to use the mass scale $M$ 
that appeared already in \eeflatmsp .  This mass scale has been interpreted as 
the characteristic inverse size of the universe (cf.\ \eedfulvol ).  This 
interpretation of $M$ is compatible with the mean field theory requirement 
that the total flux of the uniform density field $J$ be equal to that of 
the partonic current $\CJ$,  
\eqn\eeintcurr{\int_{\CM_{10}}J^0=cN.}
The multiplicative constant $c$ on the right hand side of \eecurrmfa\ and 
\eeintcurr\ is independent of $N$.  This constant measures the contribution of 
an individual parton into $J^0$, and will have to be determined a posteriori 
due to our lack of knowledge about the precise microscopic origin of 
\eecurrmfa .  

Our theory is defined by \eelagr , with ``Tr'' being the parity-invariant, 
$\osp(1|32)\times\osp(1|32)$ invariant supersymmetric extension of \eeadstr .  
Due to the presence of the current on the right-hand side of the mean field 
equations of motion, 
\eqn\eemfaeom{F\wedge\ldots\wedge F=g^2J,}
the flat eleven-dimensional space 
\eqn\eesoln{\bar A_M=M\bar e_M^AP_A}
is indeed a solution of the theory.  

When integrated over the space-like hypersurface $\CM_{10}$, the time 
component of the equations of motion requires 
\eqn\eetimeeom{\int_{\CM_{10}}F\wedge\ldots\wedge F=g^2\int_{\CM_{10}}J,}
which leads to 
\eqn\eeqc{cg^2N=1.}
We choose the value of $c$ (which is independent of $g$ and $N$) such that 
the quantized gauge coupling $k\sim 1/g^2$ is precisely equal to $N$.%
\foot{In more generality, one might consider cases with $k=mN$, with $m$ not 
necessarily equal to one (but independent of $k$ and $N$).  Assuming that the 
theory makes sense for any number of partons, $m$ has to be a positive 
integer.  In fact, this positive integer $m$ relates the number of partons 
$N$ to the size of the universe they generate, and it might be tempting to 
refer to it as the ``Mach number'' of the universe.  In this paper, we will 
only consider universes with Mach number equal to one.  This is indeed the 
most refined case -- universes with Mach number higher than one will have 
effectively less partons per given volume than the minimal case of Mach number 
one, and presumably correspond to partons in higher representations of the 
gauge group.}
In other words, the quantized gauge coupling constant $k\sim 1/g^2$ is to be 
identified with the number of partons in the system.  This resolves one of 
the puzzles about the low-energy interpretation of our theory -- the 
dimensionless gauge coupling $g$ is determined by the presence of matter 
in the system.  

\subsec{Low-energy field theory}

Now we wish to identify a regime with a well-defined low-energy effective 
description.  At first, our arguments will be independent of the precise 
supersymmetric extension of the anti-de~Sitter group.  Therefore, we will 
study the bosonic anti-de~Sitter sector of the theory first, hoping that this 
will make our arguments more transparent.  

Our theory still contains two parameters -- a mass scale $M$ introduced in 
our solution to the mean-field equations of motion, and the dimensionless 
Chern-Simons coupling that we have just identified with the number of partons 
$N$ in the system.   The requirement that the theory have a low-energy regime 
described by conventional effective theory will determine one of these 
parameters in terms of the other.  

First we rewrite the theory in terms of rescaled variables suitable for the 
anticipated low-energy supergravity regime, 
\eqn\eenewa{A_M=Me_M^AP_A+\omega_M^{AB}J_{AB}+\ldots,}
and consider the effective theory for fluctuations near the flat space-time 
solution.  Thus, we assume
\eqn\eeesmall{e_M^A-\bar e_M^A\ll 1.}

It will be convenient to replace the mean-field current $J=NM^{10}P\wedge\bar 
e\wedge\ldots\bar e$ by $NM^{10}P\wedge e\wedge\ldots\wedge e$.  This 
corresponds to an improved mean field approximation, in the following sense.  
The distribution of partons, summarized in the mean field theory by $J$, 
determines the large-scale metric in space-time; when we consider geometries 
$e$ close to but different from the flat geometry $\bar e$, the distribution 
of partons can be expected to adjust to this change of the space-time 
geometry, leading to the modified mean field expression for $J$ in which 
$\bar e$ is replaced by $e$.  Practically, this substitution allows us to keep 
general covariance in mean field theory.

The bosonic anti-de~Sitter sector of our $\osp(1|32)\times\osp(1|32)$ 
Lagrangian can be written in terms of the rescaled variables as \alicham 
\eqn\eelone{\CL=-\frac{1}{g^2}\int_\CM\sum_{s=0}^5\frac{M^{2s+1}}{2s+1}
\pmatrix{5\cr s\cr}\epsilon_{A_1\ldots A_{11}}e^{A_1}\wedge\ldots\wedge 
e^{A_{2s+1}}\wedge R^{A_{2s+2}A_{2s+3}}\wedge\ldots\wedge R^{A_{10}A_{11}}.}
($R^{AB}\equiv d\omega^{AB}+\omega^{AC}\wedge\omega_C{}^B$ denotes the 
Riemann curvature of $\omega_M^{AB}$.)

We are looking for a regime with a well defined low-energy effective 
description.  In this regime, the low-energy theory should have a kinetic 
term containing the Einstein-Hilbert term linear in $R$.  Keeping the 
Einstein-Hilbert term in \eelone\ finite, we can identify the effective 
Planck mass, 
\eqn\eeplanck{M_P\sim\frac{M}{g^{2/9}}.}

In the low-energy theory, we want to keep $M_P$ fixed.  Since $g$ is 
related to the number of partons by \eeqc , the scaling that leads to a 
well-defined low-energy theory requires $M$ to scale with the number of 
partons, such that $g\rightarrow 0$, $M\rightarrow 0$, and $Mg^{-2/9}$ is 
fixed.  Note that since $M$ is the inverse characteristic size of the 
universe, this scaling is consistent with the assumption that the universe 
is  macroscopically large in Planck units.  Note also that in terms of the 
microscopic Chern-Simons gauge theory, this regime corresponds to the 
semiclassical limit, $g\rightarrow 0$.  

We have identified the low-energy Planck length in terms of the Chern-Simons 
coupling constant $g$ and the mass parameter $M$.  Now we can look more 
closely at the low-energy effective theory.  The Lagrangian \eelone\ can be 
written in terms of $M_P$ and $g$ as follows:
\eqn\eeltwo{\eqalign{\CL&=-M_P^9\int_\CM\Tr\left(e\wedge\ldots\wedge e\wedge 
R+\frac{c_2}{g^{4/9}M_P^2}e\wedge\ldots\wedge e\wedge R\wedge R\right.\cr
&\quad\left.{}+c_0g^{4/9}M_P^2e\wedge\ldots\wedge e+\CO(g^{-8/9}M_P^{-4})
\right).\cr}}
(Here, as in \eelone , the trace is defined by the antisymmetric $\epsilon$ 
tensor; $c_0$ and $c_2$ are certain constants of order one and independent of 
$g$ and $M_P$.)

In the effective theory, we will keep only the leading term, proportional 
to $M_P^9$ and containing the term linear in curvature.  
This rule extends to the full $\osp(1|32)\times\osp(1|32)$ supersymmetric 
theory, thus leading to a low-energy supergravity with the Planck mass 
given by \eeplanck .  We have also indicated the presence of the cosmological 
constant term in the bosonic Lagrangian \eeltwo ; this term vanishes in the 
limit of infinitely large space-time, and should be absent in the full 
supersymmetric theory.  Its dependence on $g$ and $M_P$ is of some interest, 
however, and we will return to this issue briefly in Section~5.  

The effective theory that only keeps terms proportional to $M_P^9$ can only 
be valid as long as the higher-order curvature terms in \eeltwo\ are much 
smaller than the leading curvature term.  Thus, the low-energy supergravity is 
a good effective theory only at sufficiently large length scales and for 
sufficiently small space-time curvatures.  

The higher order curvature terms in \eeltwo\ are indeed suppressed by inverse 
powers of the Planck mass $M_P$.  However, powers of $g$ also appear, and we 
obtain the following condition on the space-time curvature in the effective 
theory, 
\eqn\eelecond{R^{AB}\ll g^{4/9}M_P^2.}
This is a surprisingly strong restriction on the validity of the low-energy 
effective field theory.  We will see momentarily that this should not 
be a surprise at all, as our microscopic theory turns out to satisfy the 
holographic principle.  In a holographic theory, the low-energy approximation 
by an effective field theory in large space-time suffers from a drastic 
overcounting of the number of degrees of freedom, and therefore should break 
down much before the naively expected Planckian cutoff.  The condition 
\eelecond\ is the manifestation of precisely such breakdown of the low-energy 
effective theory.  

\subsec{Low-energy symmetries: space-time diffeomorphisms}

Microscopically, our theory is a gauge theory.  We have seen that at low 
energies, the theory is effectively described by a Lagrangian linear in 
Riemann curvature.  It is known that this standard (super)gravity Lagrangian 
is not invariant under the gauge symmetries associated with translations; in 
supergravity, gauge translations are replaced by diffeomorphisms.  In our 
case, the gauge translations are clearly symmetries of our microscopic theory, 
and one may wonder how they can get replaced by diffeomorphisms in the 
effective low-energy theory.  

To see how this happens, consider the following.  
At low enough energies, the higher-curvature terms in the Lagrangian are 
small, and our theory is described to a good approximation by the low-energy 
term linear in $R$.  
The microscopic gauge symmetry algebra can be rewritten in terms of rescaled 
charges with appropriate dimensions for the low-energy theory, 
\eqn\eecharges{P_A=M^{-1}\tilde P_A,\qquad Q_\alpha=M^{-1/2}\tilde Q_\alpha.}
Schematically, the relevant part of the commutation relations is 
\eqn\eealgqq{\eqalign{\{\tilde Q,\tilde Q\}&=\Gamma^A\tilde P_A+
\frac{M_P}{N^{1/9}}\Gamma^{AB}J_{AB}+\Gamma^{A_1\ldots A_5}
K_{A_1\ldots A_5}+\ldots,\cr
[\tilde P_A,\tilde P_B]&=\frac{M_P^2}{N^{2/9}}J_{AB}+\ldots.\cr}}
(The ``$\ldots$'' refer to the higher-form charges.)  It is easy to see that 
even though this is the symmetry algebra of the microscopic theory, it is not 
a symmetry of the low-energy Lagrangian.  Indeed, under gauge translations 
$\tilde\varepsilon^A$, we have from the variation of $e_M^B$ in the effective 
Lagrangian 
\eqn\eevarlag{\delta\CL_{eff}\sim-M_P^9\int\Tr\left(\tilde\varepsilon e\wedge
\ldots\wedge e\wedge T\wedge R\right).}
(Here $T^A=de^A+\omega^A{}_B\wedge e^B$ is the torsion of $e$.)  In the 
microscopic theory, this non-invariance is canceled by the variation of a term 
which is of higher order in curvature, and gauge translations are a gauge 
symmetry.  Indeed, in the microscopic theory we have
\eqn\eevarcurv{\delta_{\tilde\varepsilon}R^{AB}\sim\frac{M_P^2}{N^{2/9}}\tilde
\varepsilon^{[A} T^{B]},}
and the variation of $R$ in the $R\wedge R$ term cancels that of \eevarlag .  
In the low-energy effective theory, however, the terms of higher order in $R$ 
are absent, and the gauge translations are not a symmetry.  Rather, the 
effective symmetry algebra of the low-energy theory is related to the 
contraction of the microscopic algebra, obtained by setting $M_P/N^{1/9}$ to 
zero in the commutation relations.  In particular, the gauge translations are 
effectively replaced in the low-energy theory by local diffeomorphisms.  

We have argued that the low-energy supergravity description breaks down as 
we reach curvatures of order $M_P^2/N^{2/9}$.  As we approach the limit 
set by \eelecond , the theory crosses over to an intermediate regime where the 
mean field approximation should still hold, since the characteristic distance 
between partons is much smaller than $N^{1/9}M_P^{-1}$.  In that regime, the 
higher curvature terms become important, and space-time diffeomorphisms are 
replaced by the microscopic gauge symmetry.  In this intermediate regime, the 
theory becomes a true gauge theory, still in a mean field approximation.%
\foot{Notice that the improved current $NM^{10}e\wedge\ldots\wedge e\wedge P$ 
is only conserved if torsion is zero.  The improved mean field theory in the 
intermediate regime where $T$ may no longer be zero would require 
modifications of the improved mean field current that take torsion into 
account.}

\subsec{Low-energy supersymmetry}

Having understood how space-time diffeomorphisms appear as a part of the 
low-energy symmetry, we now return to the full supersymmetric theory.  Our 
discussion will be brief and sketchy.  We will not try to demonstrate in 
detail whether the full low-energy theory really reproduces minimal 
eleven-dimensional supergravity of \cjs .  We will find indications suggesting 
that this should be the case, but a more detailed analysis would certainly be 
desirable.  

The full supersymmetry algebra $\osp(1|32)\times\osp(1|32)$ can be written in 
terms of the rescaled charges, 
\eqn\eefullrescf{K_{A_1\ldots A_r}=M^{-1}\tilde K_{A_1\ldots A_r},\qquad 
Q'_\alpha=M^{-3/2}\tilde Q'_\alpha.}
This rescaling is the only one compatible with that of \eecharges\ and with 
the structure of the theory at low energies.  The effective symmetry of the 
low-energy theory is related to the $M\rightarrow 0$ contraction of this 
microscopic $\osp(1|32)\times\osp(1|32)$ algebra, for reasons discussed 
briefly in the previous subsection.  

There are several arguments indicating that the low-energy theory can be 
expected to reproduce eleven-dimensional supergravity: 

(1) The low-energy symmetry algebra obtained from the contraction of the 
microscopic gauge symmetry is the algebra with 64 supercharges that was 
identified by D'Auria and Fr\'e in \hidden\ as the hidden algebra of 
eleven-dimensional supergravity.  (The extra two-form charge that appears in 
\hidden\ is to be identified with our $\epsilon_{ABC_1\ldots C_9}
\tilde K^{C_1\ldots C_9}$, while $K_{A_1\ldots A_6}$ and 
$K_{A_1\ldots A_{10}}$ decouple in the low-energy algebra.)

(2) In the previous subsection we have seen that in the low-energy theory, 
$\tilde P_A$ acts by diffeomorphisms.  Thus, the full low-energy symmetry 
group is a supersymmetric extension of the diffeomorphism group on $\CM$.  

(3) Supersymmetry of eleven-dimensional supergravity of course requires the 
presence of the abelian three-form $C$ in the low-energy spectrum.  
In the present context, $C$ appears at low energies as a composite field, or 
more precisely, as a three-form built out of the gauge field $A$.  This 
observation is not new, and was actually one of the main points of \hidden .   
More details and references on this approach to supergravity can be found 
in \refs{\morehidden,\sugrgeo}.  

$C$ is known to be odd under parity, and the explicit formula presented in 
\hidden\ that identifies $C$ as a particular composite field certainly 
satisfies this requirement.  Microscopically, there is an obvious candidate 
for $C$ in the $\osp(1|32)\times\osp(1|32)$ gauge theory: The Chern-Simons 
three-form $\omega_3(A)$ that is odd under the internal parity $\CI$.  
The microscopic Lagrangian can contain, in addition to the irreducible term 
$\omega_{11}(A)$, also Chern-Simons terms that are products of 
lower-dimensional forms,%
\foot{Up to this point, we have ignored all such factorizable Chern-Simons 
terms.  Such terms can be parity invariant and therefore can indeed appear in 
the microscopic Lagrangian.  However, for our choice of the gauge group, all 
such parity-invariant terms vanish identically if we set all 
$B_M^{A_1\ldots A_r}$ and $\eta_M$ to zero, and therefore do not affect the 
main line of arguments of this paper.} 
such as 
\eqn\eetffterm{\int_\CM\omega_3\wedge d\omega_3\wedge d\omega_3.}
In the effective theory, this term can be expected to give rise to the 
supergravity Chern-Simons term $\int C\wedge G\wedge G$, with $G\sim dC$ the 
field strength of $C$.  

It is natural to conjecture that in the low-energy supergravity regime of 
our theory, the composite field $C$ is the only field that does not decouple 
from $e_M^A,\omega_M^{AB}$ and $\psi_M^\alpha$.  Given this assumption, the 
only effective theory of the surviving low-energy degrees of freedom that 
respects all symmetries is eleven-dimensional supergravity \uniquesg .  

\newsec{Holography}

If our theory is to be a candidate for the microscopic description of 
M-theory, it should be a consistent quantum theory containing gravity.  
On very general grounds, as argued by 't~Hooft and Susskind 
\refs{\tholog-\threv}, quantum theory of gravity should be expected to 
satisfy the holographic principle.  In this section we present evidence that 
our local field theory is indeed holographic.  

We have shown above that the Chern-Simons coupling constant $g$ is identified 
via \eeqc\ with the number of partons in the system, while the mass parameter 
$M$ should be interpreted as the inverse characteristic size of the universe 
(or, more generally, the inverse characteristic size of the box that is large 
enough to enclose the system of our interest).  

Our system is made out of $N$ partons.  Its characteristic size $L$ is 
given by $M^{-1}$, which can be expressed in Planck units in terms of the 
number of partons $N$ as (in the order of magnitude) 
\eqn\eelchar{L=\frac{1}{M}=\frac{1}{g^{2/9}M_P}=\frac{N^{1/9}}{M_P}.}
In terms of the number of partons $N$ and the Planck scale $M_P$, the 
characteristic volume $\CV$ of our system is given by
\eqn\eechvol{\CV\sim L^{10}=\frac{N^{10/9}}{M_P^{10}}.}
Similarly, the characteristic area $\CA$ of the nine-dimensional surface 
surrounding our system of $N$ partons can be expressed in terms of $N$ and 
$M_P$ as follows: 
\eqn\eecharea{\CA\sim L^9=\frac{N}{M_P^9}.}
{}For the number of partons $N$ in the system we have 
\eqn\eenarea{N=\left(\frac{M_P}{M}\right)^9\sim\CA M_P^9.}
Thus, the number of partons $N$ in the system scales like the area $\CA$ of 
the nine-dimensional surface surrounding the system, measured in Planck 
units!  In precisely this sense, our theory satisfies the holographic 
principle.  

Note that in rder to derive the holographic scaling \eenarea , we have only 
used the quantization condition on the microscopic Chern-Simons coupling 
constant $g$ that relates $g$ to the number of partons in the theory, in 
combination with our requirement that the theory have a conventional 
low-energy limit described by low-energy field theory with a standard kinetic 
term.  

Having seen first indications that our theory is holographic, we can now 
return to the condition \eelecond\ that limits the domain of validity of the 
low-energy effective theory, and demonstrate that this condition is in precise 
accord with the holographic property of the theory.  In a holographic theory, 
the maximum amount of information and energy in a box of characteristic size 
$L$ should be limited by the entropy and mass of the black hole with 
Schwarzschild radius $L$ \refs{\bekenstein-\threv}.  

Consider a configuration in our theory that saturates the inequality in 
\eelecond .  This configuration carries the maximum amount of energy allowed 
for a configuration in a box of size $L$ by the condition \eelecond\ that 
expresses the bound on the validity of the low-energy effective field theory.  
In the low-energy effective theory, the energy density is given by
\eqn\eeemt{T\sim M_P^9e\wedge\ldots\wedge e\wedge R,}
and the total energy in ten-dimensional volume $\CM_{10}$ is 
\eqn\eetoten{E\sim M_P^9\int_{\CM_{10}}e\wedge\ldots\wedge e\wedge 
R.}
{}For the configuration that saturates the inequality in \eelecond , we get 
\eqn\eetotensat{E_{max}\sim M_P^9\frac{M_P^2}{N^{2/9}}\int_{\CM_{10}}e
\wedge\ldots\wedge e.}
The volume of the universe (or more generally, of the box $\CM_{10}$ that 
contains our system) is $\CV=M^{-10}$, which gives for the maximum energy 
$E_{max}$ 
\eqn\eetotentwo{E_{max}\sim\frac{M_P^{11}}{N^{2/9}}\frac{1}{M^{10}}=N^{8/9}
M_P.}
$E_{max}$ has a simple form when expressed in terms of the number of partons 
$N$ and the characteristic inverse size of the box $M$, 
\eqn\eetotenthree{E_{max}\sim NM.}
This can be further rewritten using the relation \eenarea\ between the number 
of partons $N$, the Planck mass $M_P$ and the inverse size of the box $M$, 
\eqn\eetotenfour{E_{max}=M\left(\frac{M_P}{M}\right)^9.}
This is precisely the energy of the Schwarzschild black hole with radius 
$R_S=M^{-1}$!%
\foot{This is to be contrasted with the maximum energy expected by the naive 
Planckian cutoff; indeed, configurations with curvature $R\approx M_P^2$ would 
have energy of order $M_P(M_P/M)^{10}$, i.e.\ Planckian energy per Planckian 
unit of volume.}

Thus, the low-energy effective description of the system in terms 
of conventional supergravity, as derived in the previous section, breaks down 
when the energy of the system is equal to the mass of the black hole with the 
Schwarzschild radius equal to the size $M^{-1}$ of the box surrounding the 
system -- precisely as expected in a holographic theory.  

Several remarks seem in order: 

(1) In addition to the partons represented by the Wilson lines, the 
microscopic theory contains extra degrees of freedom, in the pure Chern-Simons 
sector of the theory.  Microscopically, there will be fluctuations satisfying 
the vacuum equations of motion in the space between the Wilson line sources, 
\eqn\eecsdof{F\wedge\ldots\wedge F=0.}
Could these extra, Yang-Mills degrees of freedom spoil or modify the 
holographic property of the theory?  The answer is no, in the following 
sense.  The holographic property is a property of the low-energy supergravity 
regime.  In the mean field approximation, which is valid in large space-time 
in the supergravity regime, the extra degrees of freedom \eecsdof\ do not play 
any role -- the only low-energy degrees of freedom observable by a low-energy 
observer are those of the effective supergravity.  The theory is holographic, 
as an effective low-energy theory.%
\foot{Notice that the theory is holographic precisely to the same extent that 
it satisfies Mach's principle; macroscopic space-time geometry is determined 
by the distribution of partons alone, as long as the role of the field-theory 
degrees of freedom satisfying $F^5=0$ is negligible.}

(2) In order to describe local experiments that can be confined inside a box 
of size $L$, we can stretch the validity of the effective field theory be 
choosing $M$ as large as possible to still give enough degrees of freedom to 
describe the experiment, i.e.\ $M$ should be of order $L^{-1}$ (and not the 
inverse size of the whole universe).  In this way, the holographic property of 
the theory can be reconciled with the local validity of the low-energy field 
theory.  

(3) As we approach the regime of energies close to the bound \eelecond\ 
(which coincides, as we have seen, with the Bekenstein bound), the theory 
should cross over from the low-energy regime described by eleven-dimensional 
supergravity to an intermediate regime described by Yang-Mills gauge theory, 
still in a mean field approximation.  According to \eetotenthree , as we 
approach the limit of validity of the low-energy supergravity description, 
each parton carries energy of order $M$.  In the intermediate regime where the 
theory becomes a gauge theory in the mean field approximation, the excess 
energy will have to be carried by excited states of the individual partons, 
or by excitations of the gauge field.  

(4) The expression for $E_{max}$ can be also rewritten as 
$E_{max}=M_P(M_P/M)^8$.  This formula suggests that Planckian energy 
density is actually carried by cells of Planckian size on an 
{\it eight\/}-dimensional surface.  This is reminiscent of the intuitive 
picture in \sholog, with the system being described by some incompressible 
fluid on the holographic screen.  In this picture, the Planckian energy 
density would be carried by cells of Planckian size in the boundary of such 
incompressible fluid.  

\subsec{The cosmological constant and naturalness}

Since our field theory is a realization of the holographic principle, it might 
shed new light on the cosmological constant problem.%
\foot{The possibility that the cosmological constant problem could be solved 
in a holographic theory has been stressed repeatedly to the author by 
Tom~Banks.  See also \banks .}

Looking back at the effective theory \eeltwo\ and ignoring supersymmetry, 
we do indeed see that the cosmological constant term would be naturally 
suppressed by a negative power of the number of partons in the system, 
\eqn\eecosmoc{\Lambda\sim\frac{M_P^{11}}{N^{2/9}}.}
Of course, in the full supersymmetric theory the value of $\Lambda$ would be 
zero by supersymmetry (and uniqueness of minimal eleven-dimensional 
supergravity \uniquesg ).  We have not relied on supersymmetry in our 
arguments leading to holography, however, and we expect the arguments to hold 
in vacua with no supersymmetry, or in general, in compactifications to lower 
dimensions with $\Lambda\neq 0$. 

There are indeed two possible points of view in our theory.  On one hand, 
the low-energy field-theory observer underestimates the importance of terms of 
higher order in curvature and expects them to be suppressed by negative powers 
of $M_P$, and therefore expects the effective supergravity description to be 
valid for energies up to the Planck scale.  The same observer has a 
naturalness problem with the value of the cosmological constant \eecosmoc , 
which based on low-energy field theory alone, should be of order $M_P^{11}$.  

On the other hand, the ``microscopic'' observer who knows about the underlying 
Chern-Simons gauge theory has no problem with the small value of the 
cosmological constant, which is naturally suppressed by an inverse power of 
the number of partons.  This microscopic observer also predicts that the 
low-energy supergravity description breaks down much faster than expected by 
the low-energy observer, because the higher curvature terms (and perhaps more 
importantly, the underlying gauge invariance) become important well before 
the Planck scale.  In holographic field theory, a small cosmological constant 
seems natural.  

This argument will extend to compactifications of the theory to lower 
dimensions.  Consider for example compactifications to four-dimensions on 
a seven-manifold of volume $L^7$.  Using \eecosmoc\ and the relation 
$m_P^2=M_P^9L^7$ between the four-dimensional Planck mass $m_P$ and the 
eleven-dimensional Planck mass $M_P$ we obtain, for the four-dimensional 
energy-density, $\lambda\sim m_P^2M^2$ -- an order of magnitude estimate that 
nicely agrees with the experimental bounds on $\lambda$ \weinberg .  

\newsec{Comments}

In this paper, we have studied a local field theory in eleven dimensions, 
which contains low-energy supergravity and exhibits the holographic property 
of 't~Hooft and Susskind.  We have presented this holographic field theory 
as a possible candidate for a covariant, ``wave mechanics'' formulation of 
non-perturbative quantum M-theory.  

In this approach to M-theory, we do not suggest new ``fundamental principles'' 
for the microscopic physics at the Planck scale.  Instead, our results seem to 
support the conjecture that M-theory might be well described by an effective 
field theory, all the way to (and perhaps even beyond) the Planck scale.  Such 
effective field theory may in principle be well-defined to all energy scales 
(just as QCD is well-defined).  The expected ``low-energy'' phenomena 
(such as eleven-dimensional supergravity and the holographic principle) would 
emerge hierarchically at lower energies in this effective framework.  

We have focused our attention on the minimal theory compatible with the 
requirements of supersymmetry and parity invariance, which leads to gauge 
group $\osp(1|32)\times\osp(1|32)$ with 64 supercharges.  In the framework of 
effective theory, this minimal theory can in principle be embedded into an 
even larger theory, with bigger supersymmetry algebra.  In this respect, the 
eleven-dimensional superconformal group $\osp(1|64)$ would be a particularly 
natural choice.  Whether such an extension will be useful or necessary is 
unclear.  

Perhaps the most surprising result of this paper is the fact that 
the holographic principle is compatible with microscopic locality.  By 
microscopic locality we mean the fact that the theory is formulated in terms 
of fields (and possibly a system of partons) with a Lagrangian which is a 
local function on the underlying eleven-dimensional manifold.  Effectively, 
this microscopic locality can still lead to apparent macroscopic non-locality, 
which can manifest itself in the effective low-energy theory in effects such 
as the holographic property.  

One is naturally curious about possible relations of the holographic field 
theory to matrix theory.  We do not have much to say about this issue, 
except for noticing that it is tempting to compare the partons of the 
holographic field theory with the D0-brane degrees of freedom of matrix 
theory.  One can formulate the holographic field theory in light-cone gauge, 
and try to integrate out the gauge field degrees of freedom at low energies.  
This would leave us with an effective theory of $N$ partons, which could then 
be compared to matrix theory.  

We have studied the theory on manifolds without boundaries.  It might be 
interesting to point out that the anomaly cancellation mechanism 
\refs{\hw,\hweff} that predicts the existence of $E_8$ super Yang-Mills 
``edge states'' in M-theory on manifolds with boundaries bears a remarkable 
resemblance to the anomaly cancellation mechanism that predicts the existence 
of similar edge states in Chern-Simons gauge theory 
\refs{\csgtbound,\hallcs,\condmat,\cbound,\bala}.  

The construction presented in this paper can also be repeated in lower 
space-time dimensions $D=4p-1$, thus suggesting a possible hierarchy of 
``M-theories'' in three and seven dimensions.  The $2+1$ dimensional case is 
somewhat trivial,  but the $6+1$ dimensional case might be of more interest.  
Indeed, here we have an interesting option that does not exist in eleven 
dimensions: The gauge group can be extended to contain an extra compact group 
(say $SU(n)$), and we can try to identify regimes in which supergravity 
decouples in a flat space-time, possibly leaving only $SU(n)$ degrees of 
freedom.  

The local quantum field theory presented in this paper is described at low 
energies by supergravity, and satisfies the holographic principle of 't~Hooft 
and Susskind.  Regardless of whether or not it will play any role in our 
future understanding of M-theory, holographic field theory might be an 
interesting testing ground for questions that originally motivated the 
formulation of the holographic principle \refs{\tholog,\sholog}, most notably 
the black hole information paradox \threv .  

\bigskip\medskip\noindent
I wish to thank Tom~Banks, Itzhak~Bars, Eric~Gimon, Per~Kraus, 
Christof~Schmidhuber, John~Schwarz, Lenny~Susskind and Edward~Witten for 
useful discussions at various stages of this work.  
This work has been supported by a Sherman Fairchild Prize Fellowship, 
and by DOE grant DE-FG03-92-ER~40701.

\listrefs
\end